\documentclass[conference]{IEEEtran}
\usepackage{subfigure}
\usepackage{caption}
\usepackage{everysel}
\usepackage{keyval}
\usepackage{ragged2e}
\usepackage{cite}
\usepackage{array}
\usepackage{amsmath}
\usepackage{amsthm}
\usepackage{amssymb}
\usepackage{url}

\theoremstyle{definition}

\newtheorem{thm}{Theorem}

\newtheorem{lemma}{Lemma}

\newcommand*\xor{\mathbin{\oplus}}

\ifCLASSINFOpdf
  \usepackage[pdftex]{graphicx}
  \graphicspath{~/Documents/2010_Fall/ELEC_524/Project/figures}
  \DeclareGraphicsExtensions{.pdf}
\else
\fi

\usepackage{tikz, pgfplots}
\usetikzlibrary{arrows}
\usetikzlibrary{arrows,positioning} 
\usetikzlibrary{decorations.markings}
\tikzset{
    >=stealth',
    punkt/.style={
           rectangle,
           rounded corners,
           draw=black, very thick,
           text width=6.5em,
           minimum height=2em,
           text centered},
    pil/.style={
           ->,
           thick,
           shorten <=2pt,
           shorten >=2pt,}
    pir/.style={
           <-,
           thick,
           shorten <=2pt,
           shorten >=2pt,}
}
\definecolor{KeynoteRed}{rgb}{.678,.051, .051}
\definecolor{KeynoteBlue}{rgb}{0.008, 0.443, 0.60}
\definecolor{KeynoteLightblue}{rgb}{.635, .914, .973}
\definecolor{KeynoteYellow}{rgb}{0.859, 0.584, 0.212}
\definecolor{KeynoteYellow}{rgb}{0.859, 0.584, 0.212}
\definecolor{KeynoteSlate}{rgb}{0.239, 0.271, 0.322}
\definecolor{KeynoteGray}{rgb}{0.498, 0.529, 0.529}
\definecolor{KeynoteTextGray}{rgb}{0.325, 0.325, 0.325}
\definecolor{KeynoteLightGray}{rgb}{0.706, 0.706, 0.706}
\definecolor{KeynoteBlueGray}{rgb}{0.471, 0.533, 0.620}
\definecolor{ECEpurple}{rgb}{.169, .18, .455}
\definecolor{ECEcyan}{rgb}{.41, .62, .72}
\definecolor{ECEgray}{rgb}{.788, .827, .859}
\definecolor{ECEblueGray}{rgb}{61.2, 70.6, 70.6}
\definecolor{ECEblueGray}{rgb}{61.2, 70.6, 70.6}
\definecolor{RiceBlue}{rgb}{0, .14, .41}
\newcommand{\SIM}{structured cancellation}
\newcommand{\CSIM}{CSC}
\newcommand{\TOT}{TOT}
\newcommand{\ASIM}{ASC}
\newcommand{\RSIM}{R_{FD}^{(\mathrm{CSC})} }
\newcommand{\RSIMS}{{R_{FD}^{(\mathrm{ASC})} }}
\newcommand{\ROT}{R_{FD}^{(\mathrm{TOT})} }
\newcommand{\XOR}{\oplus}
\newcommand{\limT}{\lim_{T\rightarrow \infty}}
\newcommand{\Zplus}{\mathbb{Z}_{+}}

\begin{document}
%
\title{\vspace{29pt} Self-Interference Cancellation in Multi-hop Full-Duplex Networks via Structured Signaling}
\author{\IEEEauthorblockN{Evan Everett}
\IEEEauthorblockA{Rice University\\
Houston, TX 77005\\
\tt{evan.everett@rice.edu}}
\and
\IEEEauthorblockN{Debashis Dash}
\IEEEauthorblockA{Rice University\\
Houston, TX 77005\\
\tt{ddash@rice.edu}}
\and
\IEEEauthorblockN{Chris Dick}
\IEEEauthorblockA{Xilinx, Inc.\\
San Jose, CA 95124\\
\tt{chris.dick@xilinx.com}}
\and
\IEEEauthorblockN{Ashutosh Sabharwal}
\IEEEauthorblockA{Rice University\\
Houston, TX 77005\\
\tt{ashu@rice.edu}}

        %
}


\maketitle

\begin{abstract}
This paper discusses transmission strategies for dealing with the problem of self-interference in multi-hop wireless networks in which the nodes communicate in a full-duplex mode. An information theoretic study of the simplest such multi-hop network: the two-hop source-relay-destination network, leads to a novel transmission strategy called structured self-interference cancellation (or just ``structured cancellation'' for short). 
In the \SIM\ strategy the source restrains from transmitting on certain signal levels, and the relay structures its transmit signal such that it can learn the residual self-interference channel, and undo the self-interference, by observing the ÒportionÓ of its own transmit signal that appears at the signal levels left empty by the source. 
It is shown that in certain nontrivial regimes, the \SIM\ strategy outperforms not only half-duplex but also full-duplex schemes in which time-orthogonal training is used for estimating the residual self-interference channel.
%
%
%
%
\end{abstract}

\section{Introduction}  
\begin{figure}[htbp]
\begin{center}
	\input{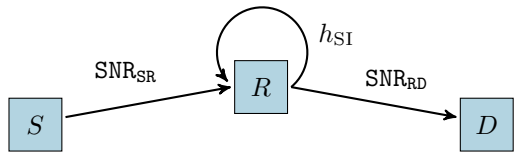}
\caption{Two-hop source-relay-destination network. The relay operates in full-duplex mode, but suffers from self-interference channel $h_{\mathrm{SI}}$. }
\label{fig:2Hop}
\end{center}
\end{figure}

Full-duplex communication can provide a significant spectral efficiency boost in multi-hop networks: relay nodes can forward packets while simultaneously receiving the next packets to be forwarded. The challenge, however, for full-duplex operation is \emph{self-interference}: a full-duplex relay's transmit signal will appear at its own receiver with very high power, potentially drowning out the signal being received.  
The two-hop source-relay-destination network depicted in Figure~\ref{fig:2Hop} is the ``unit cell'' of any multi-hop network. Transmission strategies for dealing with self-interference in the two-hop network are likely to be useful for each stage along a larger multi-hop route. As shown in Figure~\ref{fig:2Hop}, the relay operates in full-duplex mode but suffers from the presence of a self-interference channel $h_{\mathrm{SI}}$. 
 
Recent results \cite{Duarte10FullDuplex, Choi10FullDuplex,  Everett11Asilomar} have experimentally demonstrated the feasibility of full-duplex communication by employing analog and digital self-interference cancellation techniques. 
Analog self-interference cancellation is necessary to prevent the high-power self-interference from consuming the dynamic range of the A/D converter, resulting in debilitating quantization noise in the much lower power signal-of-interest.

\begin{figure}[htbp]
\begin{center}
	\input{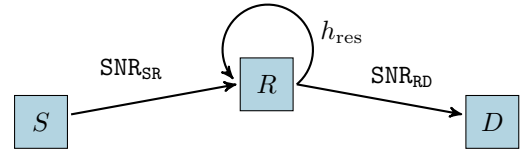}
\caption{After analog self-interference cancellation is employed, a residual self-interference channel remains.}
\label{fig:Residual2Hop}
\end{center}
\end{figure}

After analog cancellation is employed, a residual self-interference channel remains due to imperfection in the canceler, as is shown in Figure~\ref{fig:Residual2Hop}. The gain of the residual self-interference channel, $h_{\mathrm{res}}$, is unknown to the relay, for else the residual self-interference would have been subtracted off by the canceler. Moreover, $h_\mathrm{res}$ is likely to be changing with time. The analog echo cancellation technique proposed in \cite{Duarte10FullDuplex} uses a training sequence to form an estimate, $\hat h_{SI}$, of the over-the-air self-interference channel, such that the negative of the self-interference,  $-\hat h_{SI} X_R$, can be combined with the received signal. In this case the residual self-interference channel, $h_\mathrm{res} = h_{\mathrm{SI}} - \hat h_{\mathrm{SI}}$, will take on a new value every time the over-the-air self-interference is estimated. The analog echo cancellation technique proposed in \cite{Choi10FullDuplex}, uses adaptive interference cancellation, in which case the residual-self-interference channel will change as the analog echo canceler adapts. 

The results in \cite{Duarte11FullDuplex} indicate that, in some cases, due to the high power of the self-interference, residual self-interference, and not thermal noise, remains the rate-limiting bottleneck even after analog cancellation is employed. Therefore further self-interference cancellation in the digital domain is needed.

In \cite{Duarte10FullDuplex,  Everett11Asilomar} the residual self-interference is suppressed by having the source remain silent while the full-duplex relay sends a training sequence to estimate its own residual self-interference channel, so that the prediction of the residual self-interference can be subtracted off. We call this approach \emph{time-orthogonal training}. The downside to time-orthogonal training is the overhead of the training. Moreover time-orthogonal training requires no special transmission structure: the residual self-interference channel is estimated, and any self-interference left after the subtraction is treated as noise by a standard random coding scheme.

\begin{figure}[htbp]
\begin{center}
	\input{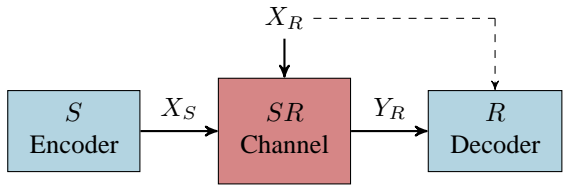}
\caption{Model of the source-to-relay link. The relay decoder has a-priori knowledge of the interfering sequence $X_R$.}
\label{fig:CodedSuppression}
\end{center}
\end{figure}

Although the relay does not know the residual self-interference channel, $h_{\mathrm{res}}$, the relay decoder has a-priori knowledge of the self-interfering sequence $X_R$ as is depicted in Figure~\ref{fig:CodedSuppression}, and can thus exploit this knowledge in signal design.
Therefore the question addressed in this paper is the following: how can we exploit knowledge of the self-interfering sequence and do better than treating residual self-interference as noise even when the residual self-interference channel is unknown and changing with time. Using the Avestimehr-Diggavi-Tse (ADT) deterministic channel model \cite{Tse11DeterministicNetworkFlow}, we study how such knowledge of the self-interfering sequence can indeed be exploited.

In this paper an information theoretic analysis of the full-duplex two-hop network suggests a structuring of the source and relay's transmit signal which we label structured self-interference cancellation (``\SIM'' for short). In the \SIM\ strategy, the relay's signal is structured such that it can learn the residual self-interference channel by observing what ``portion'' of its own data-carrying signal appears at signal levels the source has left empty for this purpose.
Once it has learned the residual self-interference channel, the relay can undo any self-interference that occurred on the signal levels carrying data from the source to the relay. In other words, instead of learning the residual self-interference channel by observing a training sequence that is time-orthogonal to the source's transmission, \SIM\ allows the relay to learn the residual self-interference channel by observing a portion of the data-carrying sequence that is signal-level-orthogonal to the source's signal. 
There are non-trivial regions in which \SIM\ achieves a higher rate than a time-orthogonal approach. In particular, \SIM\ is well-suited for situations in which the source-to-relay SNR is higher than the relay-to-destination SNR, and the residual self-interference channel coherence time is short. 

Two ``flavors'' of \SIM\ are proposed: conservative structured cancellation (\CSIM) and aggressive structured cancellation (\ASIM). The \CSIM\ scheme achieves a fixed rate for arbitrarily strong residual self-interference channels, while the \ASIM\ scheme achieves a slightly higher rate, but only when the residual self-interference is ensured to be less powerful than the signal-of-interest. The end-to-end rates that both \CSIM\ and \ASIM\ achieve in the two-hop network are derived, and performance comparisons to half-duplex, as well as to time-orthogonal training are presented.


In Section~\ref{sec:Model} the deterministic channel model for the full-duplex two-hop network is presented. Section~\ref{sec:Example} provides a motivating example that illustrates the utility of \SIM\ and is carried throughout the paper. Sections~\ref{sec:CSIM} and \ref{sec:ASIM} present the general \CSIM\ and \ASIM\ schemes, respectively, and derive the rate each scheme achieves. In Section~\ref{sec:Comparison}, the performance of the \SIM\ strategy is compared to half-duplex performance and pre-existing interference management strategies. In Section~\ref{sec:Gaussian}, a qualitative discussion on how the \SIM\ approach for the deterministic channel can be translated into a practical coding scheme for Gaussian channels. 
Concluding remarks are given in Section~\ref{sec:Conclusion}

\section{Deterministic Two-Hop Full-Duplex Channel Model}
\label{sec:Model}


Experience thus far in full-duplex communication has indicated that when a terminal operates in full-duplex mode, self-interference, not receiver noise, is the dominant rate limiting effect \cite{Duarte11FullDuplex}. Such an interference limited regime is exactly the context in which the ADT deterministic channel \cite{Tse11DeterministicNetworkFlow} is most useful. One way to think of the ADT deterministic channel model is as follows. At high SNR, Shannon's theorem tells us that for each $3$ dB increase in SNR, we get an extra bit of capacity. Therefore it makes some intuitive sense to model a channel as a set of parallel ``bit pipes'' or signal levels: one bit of information can be transmitted on each signal level per channel use. Each signal level thus corresponds to $3$ dB of above-the-noise-floor signal power. Similarly each $3$ dB of above-the-noise interference power collides with a bit's worth of the signal-of-interest. Interference is therefore modeled as an XOR operation between the bits on each of the each colliding signal levels. A more precise motivation of the ADT deterministic channel, in terms of a binary expansion of signals with noise truncating the expansion, is given in \cite{Tse11DeterministicNetworkFlow}. 

\begin{figure}[htbp]
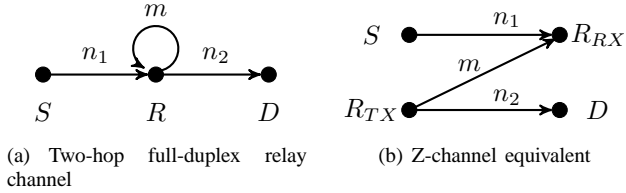

\centering
\subfigure[Two-hop full-duplex relay channel]{
	\centering
	\tikzstyle{block} = [draw,fill=KeynoteBlue!30,minimum size=2em]
\tikz[decoration={
  markings,
  mark=at position 1.0 with {\arrow{>}}}
]{
	\draw [fill=black] (1,3) circle (.1cm);
	\draw [fill=black] (2.5,3) circle (.1cm);
	\draw [fill=black] (4,3) circle (.1cm);
	\draw [->, thick] (1,3) -- (2.4,3);
	\draw [->, thick] (2.5,3) -- (3.9,3);
	\draw (1,2.5) node (S) {$S$};
	\draw (2.5,2.5) node (R) {$R$};
	\draw (4,2.5) node {$D$};
	\draw (1.7,3.25) node {$n_1$};
	\draw (3.3,3.25) node {$n_2$};
	\draw (2.5,3.85) node {$m$};
	\draw[postaction={decorate}, thick] (2.55,3.05) arc (-80:240:.3);
}
	\label{fig:chmod-multihop}
}
\subfigure[Z-channel equivalent]{
	\centering
	\tikzstyle{block} = [draw,fill=KeynoteBlue!30,minimum size=2em]
\tikz[decoration={
  markings,
  mark=at position 1.0 with {\arrow{>}}}
]{
	\draw [fill=black] (1,1) circle (.1cm);
	\draw [fill=black] (3,1) circle (.1cm);
	\draw [fill=black] (1,2) circle (.1cm);
	\draw [fill=black] (3,2) circle (.1cm);
	\draw [->, thick] (1,1) -- (2.9,1);
	\draw [->, thick] (1,2) -- (2.9,2);
	\draw [->, thick] (1,1) -- (2.95,1.95);
	\draw (0.5,2) node {$S$};
	\draw (3.52,2) node {$R_{RX}$};
	\draw (0.5,1) node {$R_{TX}$};
	\draw (3.5,1) node {$D$};
	\draw (2.3,1.17) node {$n_2$};
	\draw (2.3,2.2) node {$n_1$};
	\draw (1.8,1.6) node {$m$};
}
	\label{fig:chmod-det}
}
\caption{Multi-hop full duplex channel $(a)$ and its deterministic Z-channel equivalent $(b)$ when the relay node is thought to be split into a separate transmitter and receiver.
\label{fig:channel-model}}
\end{figure}

\begin{figure}[thbp]
\begin{center}
	\input{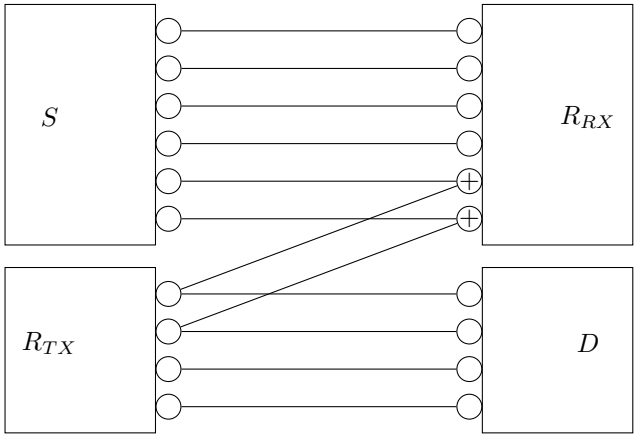}
\caption{An example of the deterministic full-duplex multi-hop channel with $n_1 = 6$, $n_2 = 4$, and $m=2$}
\label{fig:ADTexample}
\end{center}
\end{figure}

When we apply the ADT deterministic channel model to the two-hop full-duplex network of Figure~\ref{fig:Residual2Hop}, we get the model depicted in Figure \ref{fig:chmod-multihop} where $n_1 \leftrightarrow \log \mathtt{SNR_{SR}}$, $n_2 \leftrightarrow \log \mathtt{SNR_{RD}}$, and $m \leftrightarrow \log \mathtt{INR_{RR}}$, where $\mathtt{INR_{RR}} = \frac{|h_{\mathrm{res}}|^2P_R}{N_0}$ is the residual self-interference to noise ratio. 
Recognize that this model is equivalent to the well-known interference Z-channel shown in Figure~\ref{fig:chmod-det}, where the relay node is split into separate transmitter and receiver nodes. These models are equivalent in that any \emph{symmetric} rate pair achievable for the Z-channel will also be an achievable end-to-end rate in the full-duplex two-hop network. Figure~\ref{fig:ADTexample} depicts a possible instantiation of the deterministic two-hop network of Figure~\ref{fig:chmod-det}.

We assume that $n_1$ and $n_2$ are fixed and known to all terminals, but that the self-interference channel $m$ is unknown. We model $m$ as having a coherence time of $T$ channel uses. As discussed in the introduction, $m$ is unknown because it is the \emph{residual} self-interference channel left over after front-end echo cancellation is employed, and we assign $m$ a finite coherence time, because the residual self-interference channel will change as the front-end echo canceler re-estimates the over-the-air self-interference channel and adapts. 


As opposed to the classic Z-channel, we have the following situation: none of the nodes have knowledge of the cross-channel $m$, but the relay receiver, $R_{RX}$, has non-causal knowledge of the interfering bits that the relay transmitter, $R_{TX}$, is transmitting. \textbf{In other words $R_{RX}$ knows the interfering bits, but does not know at which signal levels the interference is occurring.}
Obviously, if $R_{RX}$ knew the levels at which the interference were occurring, then $R_{TX}$ could undo the interference by XOR-ing each signal level with the known interference, and interference-free communication would be the result. But can we exploit knowledge of the interfering bits without knowing which signal levels are being interfered with? In the sequel, we present an example situation in which the answer appears to be affirmative. 


\section{Motivating Example}
\label{sec:Example}
Consider an example in which $n_1 = 6$ and $n_2 = 4$.
In this case the optimal half-duplex
strategy would be to have the source-to-relay link active (all six signal levels) for $2/5$ of the time and the relay-to-destination link active for the rest of the time, which one can check gives a rate of $R_{HD} = 2.4$ bits. 
The ideal full-duplex rate (i.e. if there were no self-interference: $m=0$) is 4 bits. 
Can we exploit knowledge of the interfering sequences to outperform the HD rate and approach the ideal full-duplex rate? If so, how does finite coherence time of $m$ affect performance?  

\begin{figure}[t]
\begin{center}
	\input{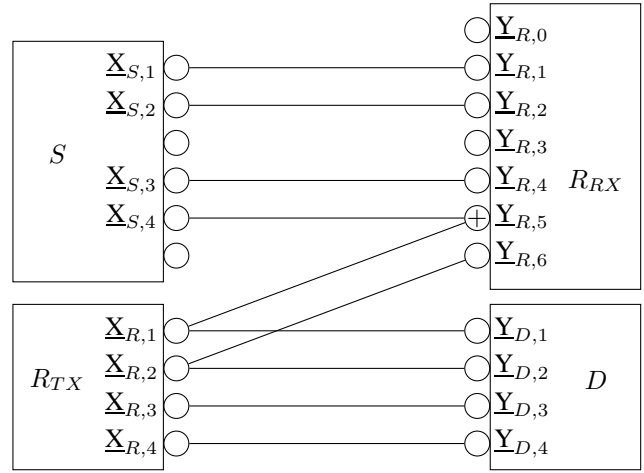}
\caption{Deterministic full-duplex multi-hop channel with $n_1 = 6$, $n_2 = 4$, and $m=2$. By observing $\underbar{Y}_{R,6} = \underbar{X}_{R,2}$, the relay decoder knows that $\underbar{Y}_{R,5} = \underbar{X}_{R,1} \xor \underbar{X}_{S,4}$, and that the other sequences from the source are interference free.}
\label{fig:example1}
\end{center}
\end{figure}

\begin{figure}[t]
\begin{center}
	\input{example2}
\caption{E $n_1 = 6$, $n_2 = 4$, and $m=6$ By observing $\underbar{Y}_{R,3} = \underbar{X}_{R,3}$, the relay decoder knows that $\underbar{Y}_{R,1} = \underbar{X}_{R,1} \xor \underbar{X}_{S,1}$, $\underbar{Y}_{R,2} = \underbar{X}_{R,2} \xor \underbar{X}_{S,2}$, $\underbar{Y}_{R,4} = \underbar{X}_{R,4} \xor \underbar{X}_{S,3}$ and $X_{S,4}$ is interference free.}
\label{fig:example2}
\end{center}
\end{figure}

Consider the transmission scheme depicted in Figure~\ref{fig:example1}, which we will call conservative structured cancellation (\CSIM). The source has six signal levels (bit pipes) at which it can transmit data, but only needs to use four levels, since the bottle-neck is the 4-level relay-to-destination link. As shown in Figure~\ref{fig:example1}, the source can restrain from transmitting on the third (middle) and sixth (bottom) levels and still achieve 4 bits per channel use. The relay transmits on all four of its signal levels, but requires that the sequences on each of the levels $\underbar{X}_{R,i},\ i\in\{1,\cdots,4\},$ be distinguishable from one another within a self-interference coherence interval $T$. This distinguishability requirement results in a source-to-relay rate less than 4 bits--the rate achieved if distinguishability were not necessary. The rate ``hit'' due to this requirement is a function of the coherence time: the shorter $T$ the worse the hit. The need for distinguishability will become apparent when the relay's decoding strategy is described. The rate limitation due to the distinguishability requirement is derived in a following section. 

In addition to the six signal levels accessible to the source, the relay receiver ($R_{RX}$) also listens to a seventh signal level: the signal level just above the highest power signal from the source (we call this the zeroth signal level). To decode, the relay first looks at $\underbar{Y}_{R,0}, \underbar{Y}_{R,3},$ and $\underbar{Y}_{R,6}$, the sequences received on the signal levels the source has left empty. From observing these sequences, the relay can infer what the interference is at each of the four signal levels being used by the source, thus allowing it to undo (via the XOR operation) any self-interference that has occurred at the data-carrying signal levels.

Figure~\ref{fig:example1} shows what would happen if $m=2$, and Figure~\ref{fig:example2} shows what would happen if $m=6$. It can be seen that for both cases (and indeed for any value of $m$) the empty signal levels reveal to the relay decoder what the interference is everywhere else, as long as the $\underbar{X}_{R,i}$'s are known to be distinguishable from each other and a null sequence.

\renewcommand{\arraystretch}{1.5}
\begin{table}[htdp]
\caption{Comparison of achievable rates for the different schemes in the $n_1 = 6$, $n_2 = 4$, example}
\begin{center}
\begin{tabular}{|l|c|c|c|c|c|}
\hline
 & $T = 1$ &$T = 2$& $T = 3$ & $T = 4$ & $T = \infty$   \tabularnewline \hline \hline
Ideal FD & 4.00 & 4.00 & 4.00 & 4.00 & 4.00  \tabularnewline \hline
HD & 2.40 & 2.40 & 2.40 & 2.40 & 2.40\tabularnewline \hline
\CSIM\ & 0 & 1.29 & 3.23 & 3.75 & 4.00   \tabularnewline \hline
\end{tabular}
\end{center}
\label{tab:ex1}
\end{table}%

It can be shown (using the the achievable rate analysis for the general \CSIM\ scheme presented in the following section) that the strategy describes above achieves that rates listed in the \CSIM\ row of Table~\ref{tab:ex1}.
We see that the \CSIM\ strategy achieves a higher rate than half-duplex as long as the self-interference remains constant for at least three channel uses. 
When $T = 3$, \CSIM\ achieves 3.23 bits, and for $T=4$ \CSIM\ gets 3.75 bits, all of which which are better than the 2.4 bit half-duplex rate. Thus even when the self-interference is changing every few channel uses, \CSIM\ still beats half-duplex and approaches the ideal full-duplex rate as $T\rightarrow \infty$.


\section{General Structured Cancellation Scheme for Unknown $m$: \CSIM}
\label{sec:CSIM}
Let us generalize the transmission and decoding strategy described in the above example, and derive the rate that \CSIM\ achieves as a function of the the channel strengths and residual self-interference coherence time: $\RSIM(n_1,n_2,T)$.

\begin{thm}
\label{res:CSIM}
Consider the deterministic full-duplex multi-hop channel of Figure~\ref{fig:channel-model} where $n_1,n_2 \in \Zplus$ are static and known, and $m\in \Zplus$ is unknown, can take on \emph{any} positive value, and has a coherence time of $T\in \Zplus$ channel uses. Under these conditions the end-to-end rate
\begin{equation}
\label{eq:RSIM}
\RSIM = \frac{1}{T} \log  \frac{(2^T-1)!}{(2^T-1-r)!},
\end{equation}
is achievable, where
\begin{equation}
r = \min((n_1-2)^+, n_2,\ 2^T-1).
\end{equation}
\end{thm}

After the transmission and decoding strategies of \CSIM\ are described in the following sections, Theorem~\ref{res:CSIM} will be derived.

\subsection{Transmission}
Here we generalize the \CSIM\ transmission scheme introduced in the previous example. In order for \CSIM\ to work, there must be two empty levels available in the source-to-relay link: one in the middle and one at the bottom. If $n_1 \geq n_2 + 2$, then we get these empty levels for free, because even while leaving the 2 levels unused on the source-to-relay link, the number of signal levels on the relay-to-destination link is still the bottleneck. However, if $n_1 < n_2 + 2$, then the number of signal levels used on \emph{both} links must be reduced to $n_1-2$  to make room for the two empty signal levels on the source-to-relay link that we need for successful decoding. \textbf{Thus in general $r = \min((n_1-2)^+, n_2)$ signal levels will be used for carrying data.} 

\begin{figure}[htbp]
\begin{center}
	\input{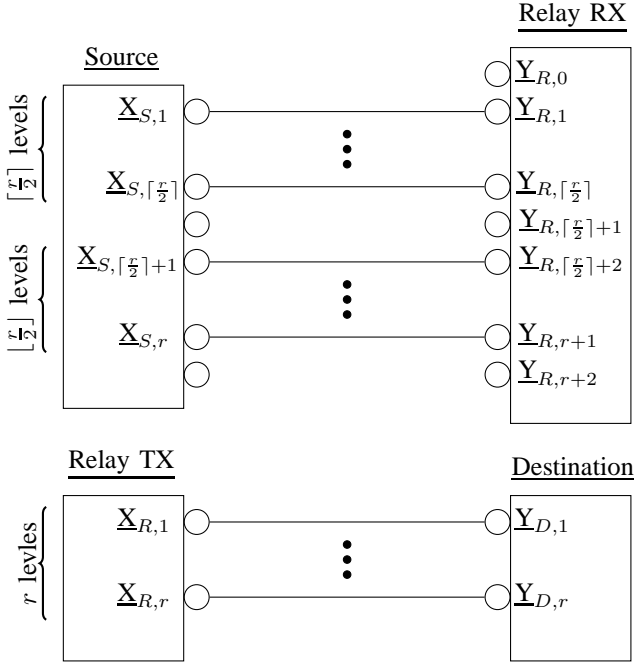}
\caption{General transmission scheme for the deterministic two-hop full-duplex network with unknown self-interference channel}
\label{fig:generalScheme}
\end{center}
\end{figure}

The general signaling scheme is depicted in Figure~\ref{fig:generalScheme}. 
The source transmits on its highest $\lceil r/2 \rceil$ signal levels, leaves a signal level open, and transmits on the next  $\lfloor r/2 \rfloor$ signal levels. Since $r \leq n_1 - 2$, this will leave at least one empty signal level at the bottom. The relay transmits on its top $r$ signal levels, but has the requirement that the sequences on each signal level are distinguishable from one another within a self-interference coherence interval $T$. There are $2^T$ such sequences. The receiver must be able to tell when a signal level is ``empty''. Because in this model receiving ``nothing'' is equivalent to receiving a sequence of all zeros, we do not allow the all-zero sequence as one of the possible sequences that may be transmitted on any given signal level. Thus there are $2^T-1$ possible sequences per coherence time $T$, and in order for the sequences on each of the $r$ signal levels to be distinguishable from one another, there must be no more than $2^T-1$ signal levels used. Hence the number of signal levels that can be used is not just limited to $\min((n_1 -2)^+, n_2)$ but $r = \min((n_1 -2)^+, n_2, 2^T-1)$.


\subsection{Decoding}
The relay decodes by first looking at the sequences received at the three empty signal levels: $\underbar{Y}_{R,0}$, the signal level just above the highest level accessible to the source, $\underbar{Y}_{R,{\lceil \frac{r}{2} \rceil}}$, the empty signal level in the middle, and  $\underbar{Y}_{R,r+2}$ the signal level just below the lowest signal level that the source uses. One can check that by observing which of the interfering sequences (or a null sequence) appear at the three empty levels, the relay can determine the interference at all signal levels being used by the source, and can thus undo (i.e. modulo subtract via XOR) the self-interference.
The destination's decoding of the relay signals is trivial.

\subsection{Limitation imposed by distinguishability requirement, and achievable rate derivation}

Let $\underbar{X}_{R,i},\ i \in \{1,\cdots,r\},$ denote the length-$T$ binary sequence transmitted by the relay on the $i$th signal level during a given coherence interval. During a length-$T$ coherence interval, there are $N=2^T-1$ unique sequences that can be chosen from. If we did not require the relay sequences to be distinguishable from one another and the null-sequnce, there would be $(N+1)^r$ unique messages per coherence interval, hence the end-to-end rate would be $1/T\log((N+1)^r) = 1/T\log(2^{Tr}) = r$, as expected. But in order for \CSIM\ to work, each of the relay sequences must be distinguishable from one another and the null-sequence. More precisely, \CSIM\ requires that 
\begin{equation}
\underbar{X}_{R,i} \neq \underbar{X}_{R,j},\ \underbar{X}_{R,i} \neq \emptyset^{(T)}\ \ \forall i\neq j\ \ i,j \in \{1,\cdots,r\}.
\end{equation}
where $\emptyset^{(T)}$ is a sequence of $T$ zeros. 


The distinguishability requirement reduces the number of possible messages that can be used. There are $N=2^T-1$ choices of different sequences for the first signal level, but only $N-1$ choices for the second signal level, and so on until the $r$th signal level, for which there are $N-(r+1)$ choices.  Thus there are $N(N-1)\cdots (N-r+1) = N!/(N-r)!$ different possible messages per coherence interval, and the achievable rate for our scheme is
\begin{equation} \nonumber
	\RSIM = \frac{1}{T} \log \left( \frac{N!}{(N-r)!} \right) = \frac{1}{T} \log \left( \frac{(2^T-1)!}{(2^T-1-r)!} \right)
\end{equation}
as is given in the theorem.


\section{General Structured Cancellation Scheme for $m < n_1$: \ASIM}
\label{sec:ASIM}
As was discussed in the introduction, practical full-duplex systems employ front-end analog cancellation prior to decoding, after which an unknown residual self-interference remains. It may be the case that the analog canceler is known to be good enough to ensure that the residual self-interference will always be weaker than the signal-of-interest. For the deterministic model under consideration, this would mean that the terminals know that $m < n_1$, although they do not know the exact value of $m$.  It turns out that such bounding of the residual self-interference allows a simpler version of the \SIM\ strategy called aggressive structured cancellation (\ASIM) that achieves a higher rate than \CSIM. In \ASIM\ the source only needs to leave one signal level empty, and the relay can undo the self-interference after observing what portion of self-interference appears at the one empty signal level. The following theorem defines the rate that \ASIM\ achieves for this $m < n_1$ situation.  

\begin{thm}
Under the same conditions as those of Theorem~\ref{res:CSIM}, with the exception that all terminals know that $m < n_1$ (but $m$ is otherwise unknown to all), the rate

\begin{equation}
\label{eq:RASIM}
\RSIMS = \frac{1}{T} \log \left( \frac{(2^T-1)!}{(2^T-1-r')!} \right)
\end{equation}
 is achievable, where
\begin{equation}
r' = \min((n_1-1)^+, n_2, 2^T-1).
\end{equation}
\end{thm}

\begin{figure}[htbp]
\begin{center}
	\input{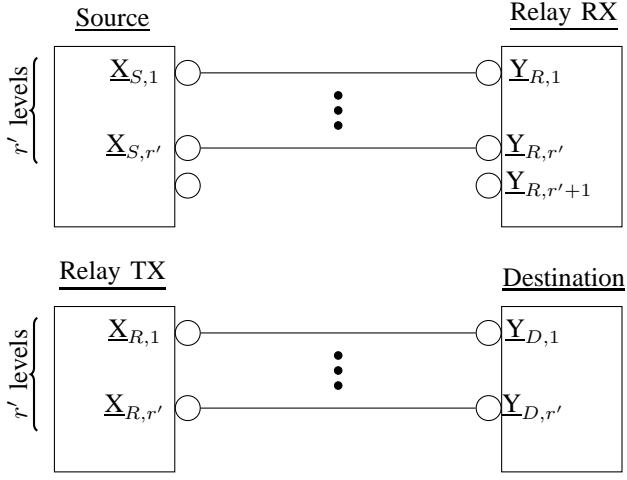}
\caption{The \ASIM\ scheme for self-interference management when $m<n_1$ is ensured.}
\label{fig:generalScheme2}
\end{center}
\end{figure}

The \ASIM\ transmission strategy for $m<n_1$ is shown in Figure~\ref{fig:generalScheme2}. The source simply transmits on its highest $r'$ signal levels. Since $r' \leq n_1 - 1$ this will leave at least one open signal level at the relay just below the signal levels the source is using. Similarly, the relay transmits on its highest $r'$ signal levels, but requires that the sequences on each signal level are distinguishable from one another within a coherence interval. 


The relay decodes by first looking at $\underbar{Y}_{R,r'+1}$ the signal level just below the source's signal. If $\underbar{Y}_{R,r'+1}$ is empty, then the relay knows that there is no self-interference: $m=0$. Otherwise $\underbar{Y}_{R,r'+1} = \underbar{X}_{R,m}$ for some $m < n_1$. From this observation the relay can determine $m$ and undo the self-interference by decoding according to 
$$\hat{\underbar{X}}_{S,i} = \underbar{Y}_{R,i},\ \ 1 \leq i < r'-m $$
and
$$ \hat{\underbar{X}}_{S,i} = \underbar{Y}_{R,i} \XOR  \underbar{{X}}_{R,(m+i)-(r'+1)},\ \ r'-m \leq i \leq r'.$$

\begin{figure}[htbp]
\begin{center}
	\input{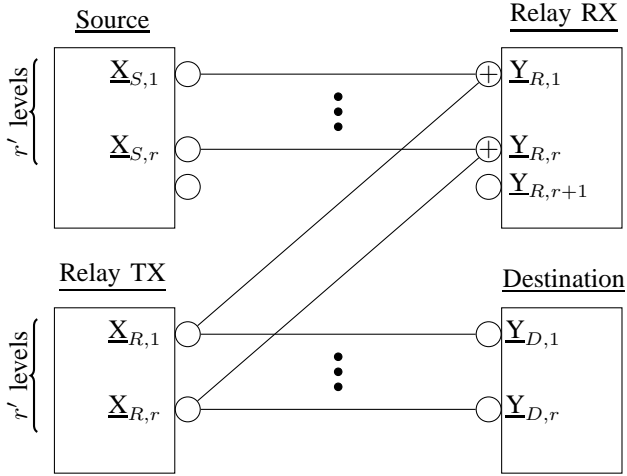}
\caption{The above illustrates why the \ASIM\ scheme only works if the relay knows $m<n_1$: when $m=n_1$ the relay cannot distinguish between this situation and a zero-interference situation.}
\label{fig:overlapExample}
\end{center}
\end{figure}

Figure~\ref{fig:overlapExample} illustrates why this particular scheme only works for $m<n_1$. If $m = n_1$ (as is shown in the figure) then the relay cannot tell by looking at the one empty signal level whether \emph{none} or  \emph{all} of the source's signal levels are being interfered with. The \CSIM\ scheme addresses this issue by putting a second empty signal level in the middle of the source's signal levels (which reduces the achievable rate) such that the interference and signal-of-interest never perfectly overlap as in Figure~\ref{fig:overlapExample}. 


\section{Performance Characterization}
\label{sec:Comparison}

\subsection{Comparison to Half-Duplex for Large Coherence Time}
The best half-duplex rate for the deterministic full-duplex two-hop channel is equivalent to the best symmetric TDMA rate for a deterministic Z-channel. Thus with a time division factor of $\alpha$, the highest half-duplex rate is $$R_{HD}=\max_{0 \leq \alpha \leq 1} \min(\alpha n_1,  (1-\alpha) n_2).$$ The rate is maximized when terms within the $\min$ are equal, i.e. $\alpha^* = \frac{n_2}{n_1+n_2}$. Therefore the best half duplex rate is given by 
\begin{equation}
R^*_{HD}=\frac{n_1n_2}{n_1+n_2}.
\label{eqn:HD}
\end{equation}

Let us begin by comparing half-duplex (HD) performance to the performance of the \CSIM\ full-duplex scheme, in the limit of long residual self-interference coherence time $T$. The following lemma gives the rate achieved by the \CSIM\ scheme is this large-$T$ regime.  
\begin{lemma}
\label{lem:limRSIM}
\begin{align}
\label{eq:limRSIM}
\limT  \RSIM = \min((n_1 -2)^+, n_2) \\
\label{eq:limRSIMS}
\limT  \RSIMS = \min((n_1 -1)^+, n_2)
\end{align}
\end{lemma}

\begin{proof}
Let $r_0 = \limT r$ be the number of signal levels used in the \CSIM\ scheme when $T$ is large. We first note that
\begin{align}
 r_0 = \limT r &= \limT \min((n_1-2)^+, n_2,\ 2^T-1) \nonumber \\
 &= \min((n_1-2)^+, n_2). \nonumber
\end{align}
Starting from Theorem~\ref{res:CSIM} we have 
\begin{align}
\lim_{T\rightarrow \infty} \RSIM &= \limT \frac{1}{T} \log \left( \frac{(2^T-1)!}{(2^T-1-r)!} \right) \nonumber \\
&= \limT \frac{1}{T} \log \left( \frac{(2^T-1)!}{(2^T-1-r_0)!} \right) \nonumber \\
 &=\limT \frac{1}{T} \log\left[ (2^T-1)(2^T-2)\cdots(2^T-r_0) \right] \nonumber \\ 
&= \sum_{k=1}^{r_0} \limT \frac{\log(2^T-k)}{T} \nonumber \\
&= \sum_{k=1}^{r_0} \limT \frac{\log(2^T)}{T} = r_0 \limT 1 = r_0 \nonumber \\
&= \min((n_1-2)^+, n_2), \nonumber 
\end{align}
which proves the first part of the lemma. The second part of the lemma easily follows by taking the same steps, but with $r'_0 = \min((n_1 -1)^+, n_2)$.
\end{proof}

%

With the help of the above lemmas, we may now specify the regime in which full-duplex \CSIM\ outperforms half-duplex (HD) in the limit of large $T$.  
\begin{thm}
\label{compare2HD}
For $T\rightarrow \infty$ and $n_1 \geq 1+\sqrt{1+2n_2}$, $$\RSIM \geq R_{HD}.$$ Conversely, for $T\rightarrow \infty$ and $n_1 < 1+\sqrt{1+2n_2}$, $$\RSIM \leq R_{HD}.$$
\end{thm}

\begin{proof}
It can be easily shown that $1 + \sqrt{2n_2 + 1} \leq n_2 + 2$ for all $n_2 \geq 0$, therefore let us split the proof into two cases: the $n_1 \geq n_2 + 2$ case and the $1 + \sqrt{2n_2 + 1} \leq n_1 < n_2 + 2$ case. 

First consider the $n_1 \geq n_2 +2 $ case, for which $\lim_{T\rightarrow \infty} \RSIM = n_2$ by Lemma~\ref{lem:limRSIM}. Since $n_2 \in \Zplus$ we can write

$$ n_2 \geq 0 \Rightarrow  n_1 + n_2 \geq n_1  \Rightarrow 1 \geq \frac{n_1}{n_1 + n_2}$$
$$ \Rightarrow n_2 \geq \frac{n_1 n_2}{n_1 + n_2} \Rightarrow \lim_{T\rightarrow \infty} \RSIM \geq R_{HD}.$$

Next consider the $1 + \sqrt{2n_2 + 1} \leq  n_1 < n_2 +2 $ case, for which $\lim_{T\rightarrow \infty} \RSIM = n_1 -2$ by Lemma~\ref{lem:limRSIM}. Starting from the hypothesis we have

$$ n_1 \geq 1+\sqrt{1+2n_2} \Rightarrow
(n_1 -1)^2 \geq  1+2n_2$$
$$ \Rightarrow n_1^2 - 2n_1 + 1 \geq 1+2n_2 \Rightarrow n_1^2 - 2(n_1 + n_2) \geq 0$$
$$\Rightarrow n_1^2 + n_1 n_2 - 2(n_1 + n_2) \geq n_1 n_2 $$
$$ \Rightarrow n_1(n_1 + n_2) - 2(n_1 + n_2) \geq n_1 n_2 $$
$$ \Rightarrow (n_1-2) \geq \frac{n_1 n_2}{n_1 + n_2}  \Rightarrow \lim_{T\rightarrow \infty} \RSIM \geq R_{HD}.$$

To show the converse, first note that if $n_1 \leq 2$, then  $\RSIM=0$, and the converse holds trivially since $R_{HD} \geq 0$. Otherwise we have $2 < n_1 < 1+\sqrt{1+2n_2}$ which implies $0 < n_1-2 < n_2$ hence $\lim_{T\rightarrow \infty} \RSIM = n_1 -2$ by Lemma~\ref{lem:limRSIM} and we can write
$$ 2 < n_1 < 1+\sqrt{1+2n_2} \Rightarrow (n_1 -1)^2 \leq 1+2n_2 $$
$$ \Rightarrow n_1^2 - 2(n_1 + n_2) \leq 0 $$
$$  \Rightarrow n_1^2 + n_1 n_2 - 2(n_1 + n_2) \leq n_1 n_2 $$
$$ \Rightarrow (n_1-2) \leq \frac{n_1 n_2}{n_1 + n_2} \Rightarrow \lim_{T\rightarrow \infty} \RSIM \leq R_{HD}.$$
\end{proof}


Theorem~\ref{compare2HD2} below is the counterpart of Theorem~\ref{compare2HD} for the \ASIM\ scheme.

\begin{thm}
\label{compare2HD2}
If all terminals know that $m<n_1$, then for $T\rightarrow \infty$ and $n_1 \geq \frac{1}{2}+\frac{1}{2}\sqrt{1+4n_2}$, $$\RSIMS \geq R_{HD}.$$
Conversely, for $T\rightarrow \infty$ and $n_1 < \frac{1}{2}+\frac{1}{2}\sqrt{1+4n_2}$, $$\RSIMS \leq R_{HD}.$$
\end{thm}

\begin{proof}
The proof is essentially the same as the proof of Theorem~\ref{compare2HD}, except the second part of Lemma~\ref{lem:limRSIM} is invoked, and the numerical values are modified accordingly. 
\end{proof}

Theorems~\ref{compare2HD} and~\ref{compare2HD2} tell us that when $n_1$, $n_2$, and $T$  are all large, full-duplex operation via \SIM\ outperforms half-duplex. However, there is another full-duplex scheme that could outperform \SIM\ in the limit of large $T$: time-orthogonal training. In  time-orthogonal training, during the first channel use of each self-interference coherence interval the source transmits a zero on each signal level, meanwhile the relay transmits a one on each of its signal levels. The relay learns $m$ by observing the highest signal level on which it receives a one instead of a zero. Knowing $m$, the relay can undo the self-interference in the remaining $T-1$ channel uses. 

The rate achieved by time-orthogonal training (TOT) is 
\begin{equation}
\label{eqn:TOT}
\ROT = \frac{T-1}{T}\min(n_1,n_2).
\end{equation} 
In the limit of large $T$, time-orthogonal training reaches the ideal full-duplex rate, $\min(n_1,n_2)$, and thus outperforms not only half-duplex but also \CSIM\ and \ASIM. For $n_1 \geq n_2 + 2$ \CSIM, \ASIM, and time-orthogonal training all achieve the ideal full-duplex rate in the limit of large $T$, but for $n_1 < n_2 + 1$ time-orthogonal training outperforms both \CSIM\ and \ASIM. The advantage of the \CSIM\ and \ASIM\ schemes, however, comes when the self-interference coherence time is finite. The \CSIM\ and \ASIM\ schemes do not require the source to ``turn-off'' while the relay learns the self-interference, and thus for finite coherence times can learn the residual self-interference channel more efficiently and outperform orthogonalized training, as will be discussed in the following section.

\subsection{Comparison of \CSIM\ to Time-Orthogonal Training for Finite Coherence Times}

Consider again the $n_1 = 6$, $n_2 = 4$ example with which we started. Table~\ref{tab:ex2} shows that, in this case, $\RSIM > \ROT\ \forall\ T > 1$. This example is somewhat favorable to \CSIM, in that it is a case in which $n_1 -2\geq n_2$, and thus the signal levels the source leaves empty in the \CSIM\ scheme comes at no cost, since the relay-to-destination is the bottleneck for both \CSIM\ and time-orthogonal training. Indeed it seems that  whenever $n_1 > n_2 + 2$ \CSIM\ is almost always preferred over time-orthogonal training.

Table \ref{tab:ex4} compares the performance of half-duplex, time-orthogonal training, and \CSIM\ for a $n_1 = 6$, $n_2 = 5$ example. In this case \CSIM\ uses 4 signal levels, while time-orthogonal training uses 5 signal levels, but sends no data in the first channel use. For very short self-interference coherence times, ($T=1,2$), half-duplex outperforms both time-orthogonal training and \CSIM. For medium-length coherence times, such as $T=3$ and $T=4$, \CSIM\ outperforms time-orthogonal training, because although it uses one less signal level than time-orthogonal training, \CSIM\ learns the residual self-interference channel in a more efficient way. However in the limit of large $T$, time-orthogonal training eventually wins out as the cost of not transmitting in the first channel use becomes negligible. 


\renewcommand{\arraystretch}{1.5}
\begin{table}[htdp]
\caption{Comparison of achievable rates for the different schemes in the $n_1 = 6$, $n_2 = 4$, example}
\begin{center}
\begin{tabular}{|l|c|c|c|c|c|}
\hline
 & $T = 1$ &$T = 2$& $T = 3$ & $T = 4$ & $T = \infty$   \tabularnewline \hline \hline
HD & 2.40 & 2.40 & 2.40 & 2.40 & 2.40\tabularnewline \hline
Ideal FD & 4.00 & 4.00 & 4.00 & 4.00 & 4.00  \tabularnewline \hline
\TOT & 0 & 2.00 & 2.67 & 3.00 & 4.00  \tabularnewline \hline
\CSIM & 0 & 1.29 & 3.23 & 3.75 & 4.00   \tabularnewline \hline
\end{tabular}
\end{center}
\label{tab:ex2}
\end{table}%


\renewcommand{\arraystretch}{1.5}
\begin{table}[htdp]
\caption{Comparison of achievable rates for the different schemes in the $n_1 = 6$, $n_2 = 5$, example}
\begin{center}
\begin{tabular}{|l|c|c|c|c|c|}
\hline
 & $T = 1$ &$T = 2$& $T = 3$ & $T = 4$ & $T = \infty$   \tabularnewline \hline \hline
HD & 2.73 & 2.73 & 2.73 & 2.73 & 2.73\tabularnewline \hline
Ideal FD & 5.00 & 5.00 & 5.00 & 5.00 & 5.00  \tabularnewline \hline
\TOT & 0 & 2.50 & 3.33 & 3.75 & 5.00  \tabularnewline \hline
\CSIM & 0 & 1.29 & 3.23 & 3.75 & 4.00   \tabularnewline \hline
\end{tabular}
\end{center}
\label{tab:ex4}
\end{table}%

\section{From Deterministic Channel to Gaussian Channel}
\label{sec:Gaussian}
Translating the \SIM\ strategy for a deterministic full-duplex two-hop channel to a corresponding coding scheme for a Gaussian full-duplex two-hop channel is an area of future work. However below is a qualitative description of what such a scheme might look like. 
The intuition gleaned from the \SIM\ approach for the deterministic full-duplex two-hop network is that by strategically leaving some ``emptiness'' in the source-to-relay signal (i.e by transmitting at slightly lower rate than capacity), and by structuring the relay signal $X_R$ such that an observation of what portion of $X_R$ appears in the empty-space, the relay can determine how the self-interference is aligned with the source's signal, and undo the self-interference.

Let's first address how the \ASIM\ scheme (the one that works only for $m<n_1$) could translate to a scheme for a Gaussian channel. In the \ASIM\ scheme for the deterministic channel, the source leaves its lowest signal level empty, and the relay decodes by observing the portion of the self-interference sequence that appears at this empty lowest signal level, from which it can infer and undo the interference at the other higher signal levels. Conventionally, (say in a MAC) the higher power (low granularity) signal is decoded first and then subtracted so that the lower power signal (fine granularity) can then be decoded. But for \ASIM\,  we need to do the opposite: the fine granularity signal must be observed before we can undo the self-interference and decode lower granularity signal. But how, in a practical Gaussian channel, do we decode a lower power ``portion'' of a signal before decoding the higher power portion? 

One possible approach would be for the relay receiver to first ``decode'' the higher power source signal together with any self-interference that may be present, not caring whether the relay is ``decoding'' the correct source message or some superposition of the source signal and higher power self-interference. This would require that the superposition of the source codeword and the relay codeword always be decodable, which suggests a structured code such as a \textbf{layered lattice code} \cite{TseBresler10Many2one, Jafar08LayeredLatticeCodes}. After the higher power signal is decoded to the nearest lattice point, the decoded lattice point is subtracted, allowing the relay decoder to observe what portion of the relay's own signal (i.e. self-interference) has appeared in the lower power ``empty space''. Assuming that we have structured the relay signal such this observation reveals what the self-interference was at higher power as well, we can then go back and undo the self-interference that conflicted with the source's signal and decode to the correct lattice point corresponding to the source message. 


%


\section{Conclusion}
\label{sec:Conclusion}
A new transmission strategy for full-duplex multi-hop terminals, called structured cancellation was proposed. Instead of re-estimating the residual self-interference channel using time-orthogonal training, the \SIM\ strategy structures the full-duplex relay's signal such that it can efficiently learn the residual self-interference channel using \emph{signal levels} left empty by the source.
Nontrivial cases were given for which full-duplex \SIM\ outperforms both half-duplex and time-orthogonal training full-duplex approaches. Although the \SIM\ strategy was designed in the context of an ADT deterministic channel model, intuitions were given for how \SIM\ could be extended to Gaussian channels using a layered lattice coding strategy. 
\cite{Everett11CPWprobe}
%

%

\ifCLASSOPTIONcaptionsoff
  \newpage
\fi

\bibliographystyle{IEEEtran}
\bibliography{IEEEabrv,/Users/evaneverett/Dropbox/Coop_Relay_Literature_Study/Bibliography/Research}

\end{document}